\documentclass[a4paper,11pt]{article}

\usepackage[normalem]{ulem}
\usepackage{color}

\usepackage{pos}

\title{Two-flavored heavy mesons' nuclear bound states}

\author*[a]{G.~N.~Zeminiani}
\author[a]{S.~L.~P.~G.~Beres}
\author[a]{K.~Tsushima}

\affiliation[a]{Laborat\'orio de F\'isica Te\'orica e Computacional (LFTC),
Programa de P\'{o}sgradua\c{c}\~{a}o em Astrof\'{i}sica e F\'{i}sica Computacional,
Universidade Cidade de S\~ao Paulo (UNICID),
  01506-000, S\~ao Paulo, SP, Brazil}

\emailAdd{guilherme.zeminiani@gmail.com}
\emailAdd{samuel.beres@hotmail.com}
\emailAdd{kazuo.tsushima@gmail.com,
kazuo.tsushima@cruzeirodosul.edu.br}

\abstract{We calculate the $B_c$- and $B_s$-nucleus bound state energies and
coordinate space radial wave functions by solving the Klein-Gordon equation
in momentum space. The attractive strong potentials for the
$B_c$ and $B_s$ mesons in nuclei are calculated from the respective mass shifts of
these mesons in nuclear matter using a local density approximation.
This negative mass shift may be regarded as a signature of partial restoration
of chiral symmetry in medium in an empirical sense, because the origin of
the negative mass shift in the present study is not
directly related to the chiral symmetry mechanism.}

\FullConference{The 21st International Conference on Hadron Spectroscopy and Structure (HADRON2025)\\
 27 - 31 March, 2025\\
Osaka University, Japan\\}

\begin{document}

\hfill {\bf LFTC-25-06/100}

\maketitle

\section{Introduction}

Hadron properties are expected to be modified in a nuclear medium.
With increasing nucleon density, the Lorentz scalar effective masses of
both the light and heavy mesons are expected to
decrease because of the partial restoration of chiral symmetry
in the light quark sector~\cite{Tsushima:2011kh}.
As a consequence of this negative effective mass shift
(attractive Lorentz scalar potential), meson-nucleus bound states
can be formed if this attractive interaction is strong enough.

We focus here on the $B^{\pm}_c$- and $B^0_s$-nucleus systems.
The mechanism that we consider for the meson interaction with the nuclear medium
is through the excitation of the intermediate state hadrons with light quarks
that appear in the lowest order one-loop self-energy of the heavy meson, namely,
by the $B$, $B^*$, $D$, $D^*$, $K$ and $K^*$ meson excitations.
The estimates of the free space and in-medium self-energies are made
employing an SU(5) effective Lagrangian density, 
neglecting any possible imaginary part of the self-energies.
The in-medium self-energies require as inputs the in-medium masses
of the intermediate-state mesons containing at least one light quark,
which are calculated by the quark-meson
coupling (QMC) model invented by Guichon~\cite{Guichon:1987jp,Saito:2005rv}.

The bound-state energies and coordinate-space radial wave functions are
obtained by solving the momentum-space Klein-Gordon (K.G.) equation for the meson-nucleus systems. 

\section{Mass shift in symmetric nuclear matter}

The in-medium mass shifts of the $B_c$ and $B_s$ mesons come from the enhancement
of the meson loop contributions to their self-energies relative to those in free space.
We consider the total $B^* D + B D^*$ contribution to the $B_c$ meson self-energy,
and the total $B^* K + B K^*$ contribution to the $B_s$ meson self-energy.

\begin{figure}[htb!]
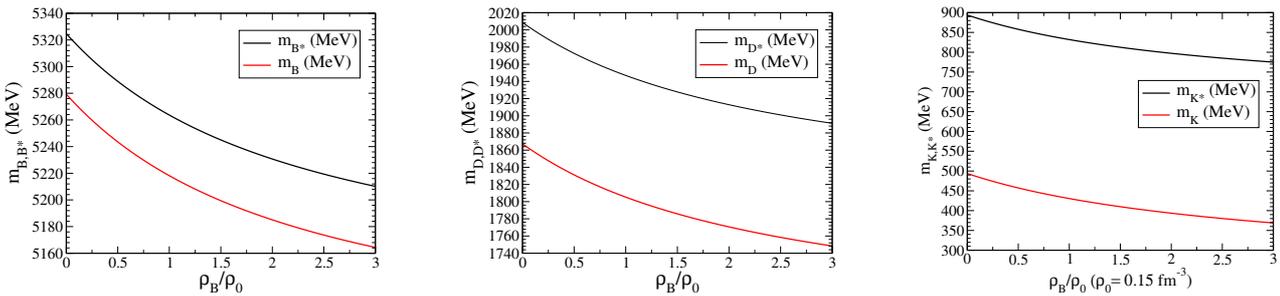
%
\centering
\hspace{-10ex}
\includegraphics[height=3.8cm]{meson_BBs_mass.eps}
\hspace{5ex}
\includegraphics[height=3.8cm]{meson_DDs_mass.eps}
\hspace{5ex}
\includegraphics[height=3.8cm]{meson_KKs_mass.eps}
\caption{$B$ and $B^*$ (left panel), $D$ and $D^*$ (middle panel), and $K$ and $K^*$
(right panel) meson Lorentz-scalar effective masses in symmetric nuclear matter
versus baryon density ($\rho_B/\rho_0$, with $\rho_0 = 0.15$ fm$^{-3}$),
calculated by the QMC
model.}%
\label{bksmass}
\end{figure}

We calculate the effective masses (Lorentz scalar) of the $B$, $B^*$, $D$, $D^*$, $K$
and $K^*$ mesons in nuclear matter as a function of nuclear matter density using the QMC model.
The results are presented in Fig.~\ref{bksmass}.
\vspace{2ex}

The mass shift is given by the difference between the meson's
(in-medium mass) - (free-space physical mass)~\cite{Zeminiani:2023gqc}.
The self-energies are calculated based on a flavor SU(5) symmetric effective Lagrangian densities,
with an SU(5) universal coupling constant value determined by the vector meson dominance (VMD)
hypothesis with the experimental data for
$\Gamma (\Upsilon \rightarrow e^+ e^-)$~\cite{Zeminiani:2020aho}.
Phenomenological form factors are used to regularize the self-energy integrals, with cutoff values
$\Lambda$
in the range 2000 MeV $\leq \Lambda \leq$ 6000 MeV.
The imaginary parts in the self-energies (corresponding to the width of the meson)
are ignored in this study. However, we plan to include the effects of the meson widths
in the near future.

\begin{figure}[htb]
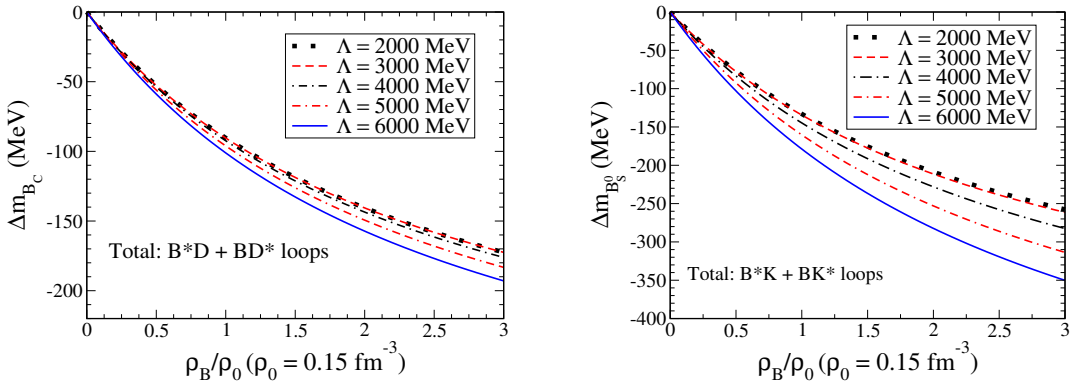
%
\centering
\includegraphics[height=5cm]{Bc_totalpot.eps}
\hspace{5ex}
\includegraphics[height=5cm]{Bs0s_totalpot.eps}
 \caption{In-medium mass shifts of the $B_c$ (left) and $B_s$ (right) mesons versus
baryon density ($\rho_B/\rho_0$) for five different
values of the cutoff mass $\Lambda$.}
\label{totbcbs}
\end{figure}

The calculated nuclear density dependent mass shifts of the $B_c$ and $B_s$ mesons are
presented in Fig.~\ref{totbcbs} as a function of the nuclear matter density ($\rho_B/\rho_0$)
for the different values of the cutoff mass $\Lambda$.
The $B_c$ mass shift $\Delta m_{B_c}(BD^*+B^*D) \equiv m^*_{B_c} - m_{B_c}$,
ranges from -90.4 to -101.1 MeV at the saturation density ($\rho_0 = 0.15$ fm$^{-3}$).
For the $B_s$ meson, the mass shift
$\Delta m_{B_s}(B^*K + BK^*) \equiv m^*_{B_s} - m_{B_s}$, ranges from
-133.0 to -178.8 MeV at $\rho_0$.
\vspace{2ex}

\section{$B^{\pm}_c$- and $B^0_s$-nucleus bound states}

We follow the procedure adopted in~\cite{Zeminiani:2024dyo}
and consider the $B_c$ and $B_s$ mesons produced inside a nucleus $A$ 
(= $^4$He, $^{12}$C, $^{16}$O, $^{40}$Ca, $^{48}$Ca, $^{90}$Zr, $^{208}$Pb)
assuming the situation in which the relative momentum of the meson to the nucleus is close to zero.
The nuclear potentials for the meson-nucleus systems
are calculated using a local density approximation, by mapping the density-dependent
meson mass shift to the nuclear density distribution of the nucleus $A$, $\rho^{A}_{B} (r)$,
where $r$ is the distance from the center of the nucleus.
The density profiles of the nuclei are calculated by the QMC model~\cite{Saito:1996sf},
except for the $^4$He nucleus, for which we use the parameterized nuclear density
from Ref.~\cite{Saito:1997ae}.

For the $B^{\pm}_c$-nucleus systems, we have to consider the Coulomb interaction.
The Coulomb potentials in the nuclei (nucleon density distribution and Coulomb mean field), with
the $B^{\pm}_c$ mesons absent, are also self-consistently obtained within the QMC
model~\cite{Saito:1996sf}. We assume that the feedback of the Coulomb force from $B_c^{\pm}$ in the
nucleus is negligible.
In this study we do not include the $B_c^{\pm}$-$^{4}$He systems, because the mean-field
approximation of the QMC model is not good for a light nucleus such as the $^4$He.
(However, we will present the $B_s$-$^4$He results only to give some ideas.)

\begin{figure}[htb]
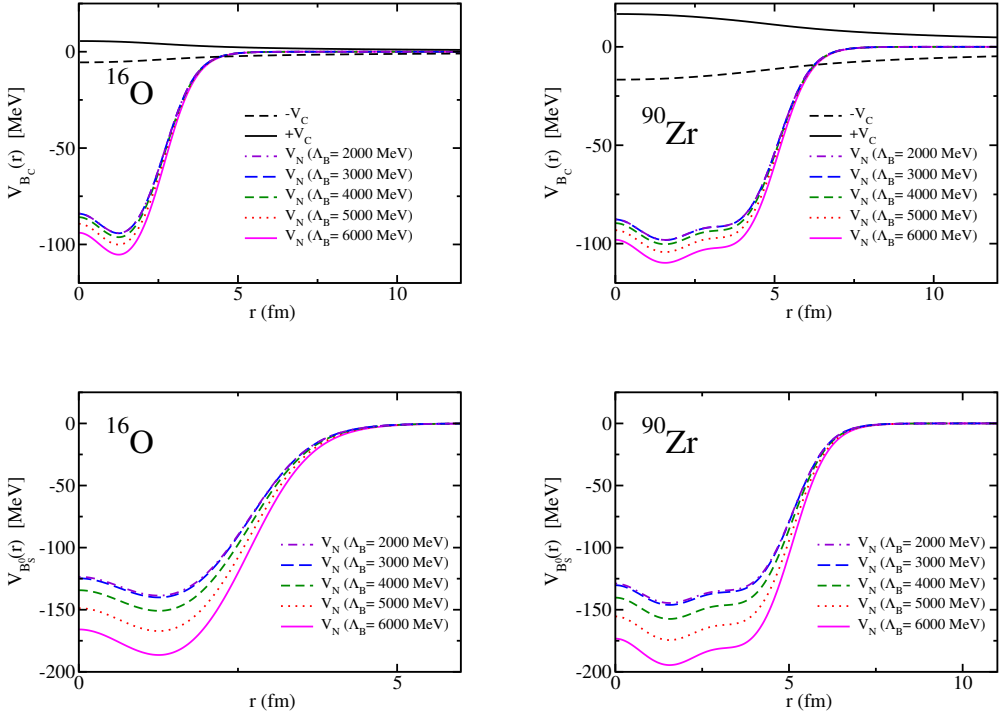
%
\centering
\includegraphics[width=6cm]{Bc_nucl_Coul_pot_O16.eps}
\hspace{5ex}
\includegraphics[width=6cm]{Bc_nucl_Coul_pot_Zr90.eps}
\\
\vspace{5ex}
\centering
\includegraphics[width=6cm]{Bs_nucl_pot_O16.eps}
\hspace{5ex}
\includegraphics[width=6cm]{Bs_nucl_pot_Zr90.eps}
 \caption{Nuclear (Lorentz scalar) and Coulomb potentials for the
$B^{\pm}_c$-$^{16}$O (top-left) and
$B^{\pm}_c$-$^{90}$Zr (top-right),
and the nuclear potentials for the
$B^0_s$-$^{16}$O (bottom-left) 
and $B^0_s$-$^{90}$Zr (bottom-right)
systems for different values of the cutoff parameter
$\Lambda$.}
\label{nuclpots}
\end{figure}

In Fig~\ref{nuclpots} we present the nuclear and Coulomb potentials for the $B^{\pm}_c$-$^{16}$O
and $B^{\pm}_c$-$^{90}$Zr systems, as well as the nuclear potentials for the $B^0_s$-$^{16}$O and
$B^0_s$-$^{90}$Zr systems.
\vspace{1ex}

\begin{figure}[htb]
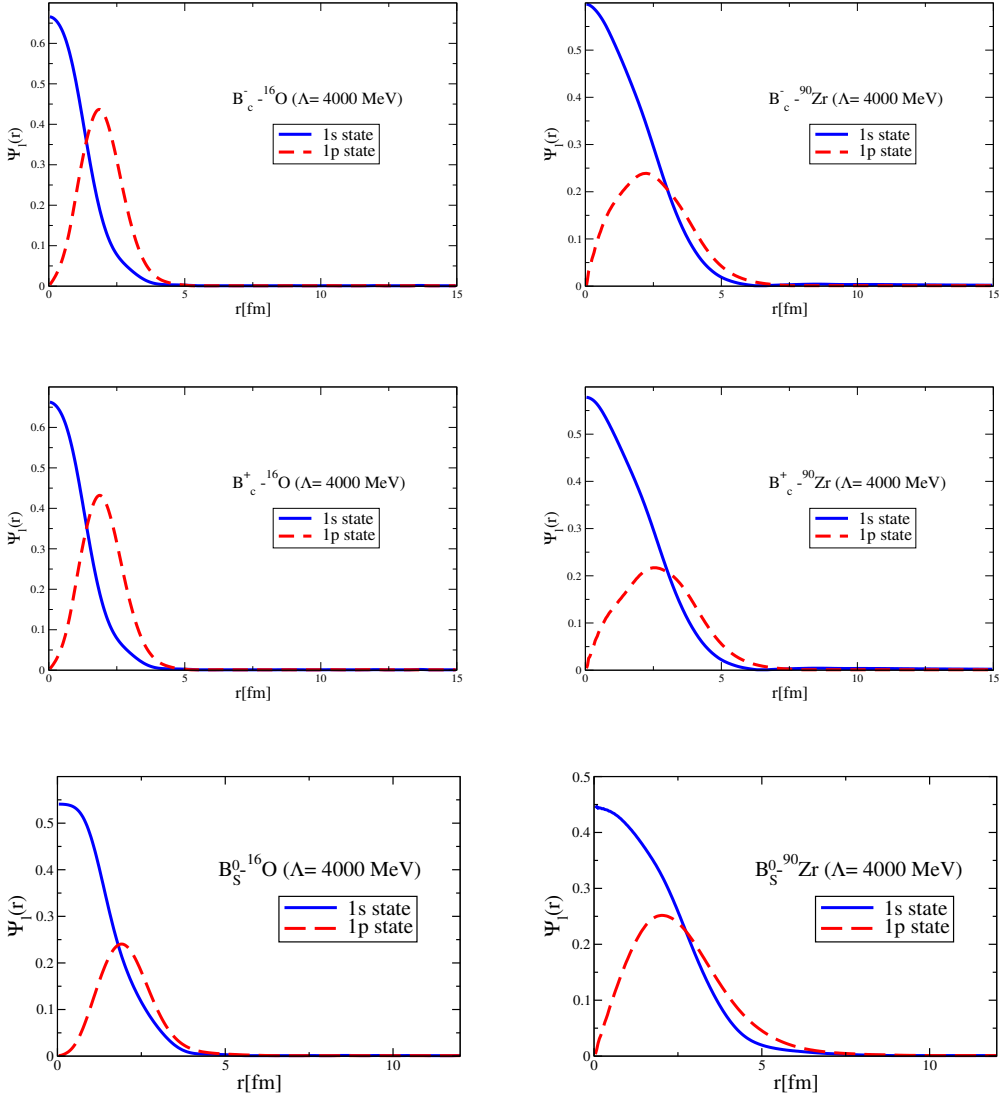
%
\centering
\includegraphics[width=6cm]{Psir_Bc_AttCoul_O16_4000_1s1p.eps}
\hspace{5ex}
\includegraphics[width=6cm]{Psir_Bc_AttCoul_Zr90_4000_1s1p.eps}
\\
\vspace{5ex}
\centering
\includegraphics[width=6cm]{Psir_Bc_RepCoul_O16_4000_1s1p.eps}
\hspace{5ex}
\includegraphics[width=6cm]{Psir_Bc_RepCoul_Zr90_4000_1s1p.eps}
\\
\vspace{5ex}
\centering
\includegraphics[width=6cm]{Psir_Bs_O16_4000_total.eps}
\hspace{5ex}
\includegraphics[width=6cm]{Psir_Bs_Zr90_4000_total.eps}
 \caption{Coordinate-space wave functions for the $B^-_c$-$^{16}$O (top left), $B^-_c$-$^{90}$Zr (top right),
 $B^+_c$-$^{16}$O (center left), $B^+_c$-$^{90}$Zr (center right),
$B^0_s$-$^{16}$O (bottom left) and $B^0_s$-$^{90}$Zr (bottom right) systems
for the central value of the cutoff parameter $\Lambda = 4000$ MeV.}
\label{wvf}
\end{figure}

We numerically solve the K.G. equation in momentum space for each value of
the angular momentum $\ell$ of the meson-nucleus system to calculate the bound-state energies of
each energy level.
The partial wave solutions for the bound-state energies and the corresponding momentum-space wave
functions of the K.G equation require the decomposition of the nuclear and Coulomb potentials into
angular momentum $\ell$ waves.
We apply the double spherical Bessel transform on the coordinate-space potentials to obtain the
$\ell$-wave decomposition
of the potentials in momentum space. The results for the bound-state energies of the $B^{\pm}_c$-A
and $B^{0}_s$-A systems are respectively presented in Tables~\ref{tblbc} and~\ref{tblbs}.
Each converged eigenvalue is then confirmed by analyzing the number of nodes in the corresponding
coordinate-space wave function, which is obtained by a spherical Bessel transform from the
normalized wave function in momentum space. We present the coordinate-space wave functions for the
$B^{\pm}_c$-$^{16}$O, $B^{\pm}_c$-$^{90}$Zr, $B^0_s$-$^{16}$O and
$B^0_s$-$^{90}$Zr systems for the central value of the cutoff parameter $\Lambda = 4000$ MeV in
Fig.~\ref{wvf}.

\begin{table}
\caption{\label{tblbc} $B^{\pm}_c$-A
bound-state energies.
The $\Lambda$ values are in MeV. }
\begin{center}
\begin{tabular}{ll|r|r|r||ll|r|r|r}
  \hline \hline
  & & \multicolumn{8}{c}{Bound state energies (MeV)}  \\
  \hline
  & & \multicolumn{2}{c}{$B^-_c$-A} &  & & & \multicolumn{3}{c}{$B^+_c$-A}\\
\hline
& $n\ell$ & $\Lambda=2000$ & $\Lambda= 4000$ &
$\Lambda= 6000$
& & $n\ell$ & $\Lambda=2000$ & $\Lambda= 4000$ &
$\Lambda= 6000$\\
\hline
$^{12}_{B^{-}_c}\text{C}$
& 1s & -79.12 & -80.63 & -87.03 
& $^{12}_{B^{+}_c}\text{C}$
& 1s & -71.01 & -72.53 & -78.94 \\
& 1p & -56.15 & -57.53 & -63.38 
& & 1p & -49.11 & -50.49 & -56.32 \\
\hline
$^{16}_{B^{-}_c}\text{O}$
& 1s & -75.00 & -76.16 & -80.94 
& $^{16}_{B^{+}_c}\text{O}$
& 1s & -64.64 & -65.80 & -70.59 \\
& 1p & -54.86 & -56.13 & -61.55 
& & 1p & -45.94 & -47.22 & -52.64 \\
\hline
$^{40}_{B^{-}_c}\text{Ca}$
& 1s & -104.27 & -105.69 & -111.87 
& $^{40}_{B^{+}_c}\text{Ca}$
& 1s & -84.89 & -86.31 & -92.49 \\
& 1p & -81.71 & -83.51 & -91.34 
& & 1p & -62.80 & -64.57 & -72.23 \\
\hline
$^{48}_{B^{-}_c}\text{Ca}$
& 1s & -96.63 & -98.37 & -105.81 
& $^{48}_{B^{+}_c}\text{Ca}$
& 1s & -77.09 & -78.83 & -86.26 \\
& 1p & -72.02 & -73.56 & -80.24 
& & 1p & -53.64 & -55.13 & -61.60 \\
\hline
$^{90}_{B^{-}_c}\text{Zr}$
& 1s & -96.34 & -98.32 & -106.82 
& $^{90}_{B^{+}_c}\text{Zr}$
& 1s & -65.51 & -67.49 & -75.99 \\
& 1p & -83.82 & -85.44 & -92.35 
& & 1p & -55.11 & -56.75 & -63.75 \\
\hline
$^{208}_{B^{-}_c}\text{Pb}$
& 1s & -95.88 & -97.39 & -103.79 
& $^{208}_{B^{+}_c}\text{Pb}$
& 1s & -48.61 & -50.13 & -56.53 \\
& 1p & -70.46 & -71.76 & -77.34 
& & 1p & -29.27 & -30.58 & -36.22 \\
\hline
\end{tabular}
\end{center}
\end{table}

\begin{table}
\caption{\label{tblbs} $B^{0}_s$-A
bound-state energies. The $\Lambda_B$ values are in MeV. }
\begin{center}
\begin{tabular}{ll|r|r|r}
  \hline \hline
  & & \multicolumn{3}{c}{Bound state energies (MeV)} \\
\hline
& & $\Lambda_{B}=2000$ MeV& $\Lambda_{B}= 4000$ MeV&
$\Lambda_{B}= 6000$ MeV\\
\hline
$^{4}_{B_s}\text{He}$
& 1s & -139.31 & -153.09 & -192.81 \\
& 1p & -111.39 & -121.60 & -150.64 \\
\hline
$^{12}_{B_s}\text{C}$
& 1s & -102.42 & -110.03 & -131.56 \\
& 1p & -80.02 & -87.85 & -111.58 \\
\hline
$^{16}_{B_s}\text{O}$
& 1s & -96.67 & -102.63 & -117.88 \\
& 1p & -74.93 & -82.25 & -103.97 \\
\hline
$^{40}_{B_s}\text{Ca}$
& 1s & -88.57 & -96.49 & -118.95 \\
& 1p & -79.59 & -88.26 & -113.66 \\
\hline
$^{48}_{B_s}\text{Ca}$
& 1s & -99.47 & -108.47 & -133.64 \\
& 1p & -86.24 & -94.89 & -119.81 \\
\hline
$^{90}_{B_s}\text{Zr}$
& 1s & -114.09 & -124.75 & -155.42 \\
& 1p & -74.69 & -82.73 & -104.76 \\
\hline
$^{208}_{B_s}\text{Pb}$
& 1s & -122.08 & -133.47 & -166.62 \\
& 1p & -100.31 & -110.50 & -140.54 \\
\hline
\end{tabular}
\end{center}
\end{table}

\section{Summary and Conclusion}

We have calculated the bound-state energies of the $B^{\pm}_c$- and $B^0_s$-nucleus systems
by solving the Klein-Gordon equation in momentum space, with the attractive Lorentz scalar
potentials originating from the mesons' mass shifts in symmetric nuclear matter.
The results indicate that the attractive nuclear potentials are strong enough to bind the
$B_c$ and $B_s$ mesons to atomic nuclei, although we have neglected the widths and the
momenta of the mesons. We plan to include these effects in the near future.

\begin{acknowledgments}
G.N.Z. and S.L.P.G.B~were supported by the Coordena\c{c}\~ao de Aperfei\c{c}oamento de Pessoal
de N\'ivel Superior-Brazil (CAPES), and FAPESP Process No.~2023/07313-6.
The work of K.T. was
supported by Conselho Nacional de Desenvolvimento
Cient\'{i}ıfico e Tecnol\'{o}gico (CNPq, Brazil), Processes
No. 304199/2022-2, and~FAPESP No. 2023/07313-6.
This work was also part of the projects, Instituto Nacional de
Ci\^{e}ncia e Tecnologia - Nuclear Physics and Applications
(INCT-FNA), Brazil, Process No. 464898/2014-5.
\end{acknowledgments}

\end{document}